\documentclass[aps,prl,twocolumn,superscriptaddress,amsmath,amsfonts,
               floatfix]{revtex4-1}

\usepackage{epsfig,psfrag}
\usepackage{CJK}

\usepackage{graphicx}
\usepackage{dcolumn}
\usepackage{bm}
\usepackage{amssymb}

\begin{document}
\title{Anharmonic corrections to the multiphonon deep-level charge capture \emph{ab initio} calculations for semiconductors}
\author{Yao Xiao$^\ddagger$}
\affiliation{State Key Laboratory of Superlattices and Microstructures, Institute of Semiconductors, Chinese Academy of Sciences, Beijing 100083, China}
\affiliation{Tianjin Key Laboratory of Low Dimensional Materials Physics and Preparing Technology, Department of Physics, School of Science, Tianjin University, Tianjin 300354, China}
\author{Zi-Wu Wang$^\ddagger$}
\affiliation{Tianjin Key Laboratory of Low Dimensional Materials Physics and Preparing Technology, Department of Physics, School of Science, Tianjin University, Tianjin 300354, China}
\author{Lin Shi}
\affiliation{Suzhou Institute of Nano-Tech and Nano-Bionics, Chinese Academy of Sciences, Suzhou 215123, China}
\author{Xiangwei Jiang$^\star$}
\affiliation{State Key Laboratory of Superlattices and Microstructures, Institute of Semiconductors, Chinese Academy of Sciences, Beijing 100083, China}
\author{Shu-Shen Li}
\affiliation{State Key Laboratory of Superlattices and Microstructures, Institute of Semiconductors, Chinese Academy of Sciences, Beijing 100083, China}
\author{Lin-Wang Wang$^\dag$}
\affiliation{Materials Science Division, Lawrence Berkeley National Laboratory, Berkeley, California 94720, USA}

\email{Email addresses: $^\star$xwjiang@semi.ac.cn; $^\dagger$lwwang@lbl.gov}

\begin{abstract}
Nonradiative carrier recombination at semiconductor deep centers is of great importance to both fundamental physics and device engineering. In this letter, we provide a revised analysis of \emph{K. Huang}'s original nonradiative multi-phonon (NMP) theory with \emph{ab initio} calculations. First, we identify at first-principle level that \emph{Huang}'s concise formula gives the same results as the matrix based formula, and \emph{Huang}'s high temperature formula provides an analytical expression for the coupling constant in Marcus theory. Secondly, the anharmonic effects are corrected by taking into account local phonon mode variation at different charge states of the defect. The corrected capture rates for defects in GaN and SiC agree well with experiments.
\end{abstract}

\maketitle
 Nonradiative transitions in semiconductors associated with impurities and defects plays a predominate role in determining many fundamental properties of semiconductors. It has been a long-sought goal in this field to theoretically predict the nonradiative decay rates of different defects. Accurate theoretical prediction is particularly important given the fact that it is very difficult to experimentally probe such transition rate for a given defect. The nonradiative multi-phonon (NMP) transition has first been theoretically studied by \emph{S. K. Pekar} \cite{wl1} and \emph{K. Huang} \cite{wl2} in 1950s. Extensive theoretical studies in this topic have been carried out over the years \cite{wl3,wl4,wl5,wl6,wl7,wl8,wl9,wl10,wl11,wl12,wl13}.

 Although there are already many theoretical studies in molecules and organic systems \cite{wl14,wl15,wl16}, only recently, \emph{ab initio} NMP calculations have been realized for defects in semiconductors. One issue is the high cost of calculating all the electron-phonon coupling constants for a large supercell containing one point-defect. A new variational principle algorithm has removed this hurdle \cite{wl17}. Since then, there are many studies using \emph{ab initio} methods to calculate the nonradiative decay rates for various systems \cite{wl18,wl19,wl20,wl21,wl22}. This, however, is not without problems. The currently used formulas all assume simple harmonic approximation for the phonon modes. This includes the harmonic phonon modes themselves at different defect charge states, as well as the change of phonon modes before and after the electron transition. For example, with a few exceptions for  molecular systems \cite{wl23,wl24}, all the current formalisms are based on the assumption that the phonon modes before and after the electron transition are the same, only their equilibrium ground state positions have been shifted. This by itself is a manifestation of a harmonic approximation in the Hamiltonian (the zero-point shift is caused by a diagonal electron-phonon coupling term in the Hamiltonian). Unfortunately, for most of the real systems, these are far from the truth. There are attempts to overcome this problem, \emph{e.g.}, by calculating the energies along the 1D reaction coordinates before and after the electron transition \cite{wl25}. But as we pointed out \cite{wl18}, such 1D method might miss the true electron-phonon coupling nuclear displacements. It is thus desirable to have a practical yet accurate procedure to take into account such anharmonic effects. This is the main purpose of the current work.

We proceed by first following \emph{Huang}'s derivation in 1980s \cite{wl8,wl26}. After a long debate between using the adiabatic coupling and static coupling approximations, \emph{Huang} has derived a formula for the static coupling NMP. In the static coupling, the electron transition happens between the adiabatic eigen states at a given (fixed) nuclear coordination. It is the nuclear movement (phonon) that introduces the perturbation, hence causes the transitions between these adiabatic eigen states. \emph{Huang} arrived at the following expression (supplemental information SI-I):
\begin{equation}
\begin{split}
&W_{ij}=\frac{2\pi}{\hbar}\!\int_{-\infty}^{\infty}\{[\!\sum_k\Omega_k^{ij}Q_{jik}(\!\cos(\eta\hbar\omega_k)+\!i\coth(\frac{\beta\hbar\omega_k}{2})\\
&\sin(\eta\hbar\omega_k))]^2+\frac{1}{2}\sum_k|\Omega_k^{ij}|^2(\frac{\hbar}{\omega_k})(\coth(\frac{\eta\hbar\omega_k}{2})\cos(\eta\hbar\omega_k)\\
&+i\sin(\eta\hbar\omega_k))\}(\frac{1}{2\pi})\exp[-i\eta\Delta E_{ji}-\sum_k(\frac{\omega_k}{2\hbar})Q_{jik}^2\\
&(\coth(\frac{\beta\hbar\omega_k}{2})(1-\cos(\eta\hbar\omega_k))-i\sin(\eta\hbar\omega_k))]d\eta,
\end{split}
\end{equation}
where W$_{ij}$ and $\Omega_k^{ij}=\langle{i}|\partial{H}/\partial{Q_k}|{j}\rangle$ are the transition rates and electron-phonon coupling constants between initial electronic states $i$ and final electronic state $j$, respectively. $k$ is the index of phonon modes with frequency $\omega_k$, ${\Delta E_{ji}}={E_i^0}-{E_j^0}$ is the transition energy of these two electronic states at their equilibrium nuclear positions $Q_{ik}^0$ and $Q_{jk}^0$. ${Q_{jik}}={Q_{jk}^0}-{Q_{ik}^0}$ denotes the phonon mode equilibrium point displacements between the initial and final states. This can also be calculated via ${Q_{jik}}=\Sigma_R{M_R\mu_k(R)\Delta R_{ji}}$, where $\Delta R_{ji}$ are the equilibrium point atomic displacements from electron state $i$ to $j$, $\mu_k(R)$ is the \emph{k}th phonon mode, $M_R$ is the nuclear mass for atom at \emph{R}. In the sum over\emph{ R}, we have used \emph{R} to indicate the three coordinates for each atom.

Furthermore, $\Sigma_R{M_R\mu_k(R)\mu_{k'}^*(R)}=\delta_{k,{k'}}$ and for a given set of phonon mode amplitudes $Q_k$, we have the atomic displacements as $\Delta R(R)=\Sigma_k{Q_k\mu_k(R)}$. The Eq.(1) is simple and straight forward to execute. On the other hand, in our previous work, we have used the formalism of \emph{Borrelli} \emph{et al.} \cite{wl28} where matrix manipulation was used, resulting in a very different looking, and much more complicated analytical expression. Thus, as a cross check, it will be interesting to compare the results of these two formalisms.

More importantly, the concise formalism of Eq.(1) allows \emph{Huang} \cite{wl26} to apply the steepest decent approximation to the exponent to obtain a close analytical formula in the high temperature limit (which will be called \emph{Huang}'s high-T formula in the following, SI-II):
\begin{equation}
W_{ij}\approx\frac{1}{\hbar}\left(\frac{\pi k_B T}{S\overline{\hbar\omega_k}}\right)^{\frac{1}{2}}\sum_k\left(\frac{\Omega_k^{ij}}{\omega_k}\right)^2e^{-\frac{(\Delta E_{ji}-S\overline{\hbar\omega_k})^2}{4k_BTS\overline{\hbar\omega_k}}},
\end{equation}
where $S=\Sigma_k{(\omega_k/{2\hbar})Q_{jik}^2}$ is the well-known \emph{Huang-Rhys} factor and ${S\overline{\hbar\omega_k}}={\Sigma_k{({\omega_k^2}/2)Q_{jik}^2}}=\lambda$ is the reorganization energy under harmonic approximation. The electron-phonon coupling constant $\Omega_k^{ij}$ can be expressed as $\langle{\psi_i}|\partial{H}/\partial{Q_k}|{\psi_j}\rangle=\Sigma_R{\mu_k(R)\langle{\psi_i}|\partial{H}/\partial{R}|{\psi_j}\rangle}$. The above formula can be compared with Marcus theory \cite{wl29} where the $(\frac{\Omega_k^{ij}}{\omega_k})^2$ is replaced by $\frac{|V_c|^2}{k_BT}$ with $V_c$ being the coupling constant (see Eq.S25). It is thus clear that \emph{Huang}'s formula provides an explicit expression for the coupling constant in Marcus theory, where its meaning and calculation is very often a subject of debate. \emph{Huang}'s formula show how that coupling constant can be expressed by electron-phonon couplig at least for phonon induced coupling case.

In the following, we chose GaN \cite{wl31,wl32,wl33} and SiC \cite{wl34}as our testing cases because there are well tested defects in these systems with known experimental nonradiative decay rates (see SI). We will use GaN:C$_N$+O$_N$ as our main example system, while present the details of the other systems in SI.

Before hole transition, the GaN:C$_N$+O$_N$ center is neutral and fully occupied. After a hole transition from the valence band, the defect is in its positive charge state (GaN:C$_N$+O$_N$$)^+$. To obtain the NMP transition rate from valence band hole to defect state using Eq. (1) and (2), the defect 0/+ transition energy, the reorganization energy, atomic displacement $\triangle{R_{ji}}$ are calculated all under the screened hybrid functional of HSE. The phonon modes $\mu_k(R)$ and its eigen frequency $\omega_k$ are calculated using the mix-dynamic matrix method as we introduced in our previous study\cite{wl17}. We have compared HSE and GGA phonon modes for both bulk and impurity systems, and found that after a simple scaling of phonon frequency they agrees very well. The electron-phonon coupling constants $\langle{\psi_i}|\partial{H}/\partial{R}|{\psi_j}\rangle$ are calculated under GGA, but with HSE wave functions. This is justified as we do not expect strong difference caused by explicit exchange term since it has no explicit \emph{R}-dependence.

\begin{figure}[htbp]
\centering
\includegraphics[bb=0 0 1280 1000,width=3.8in]{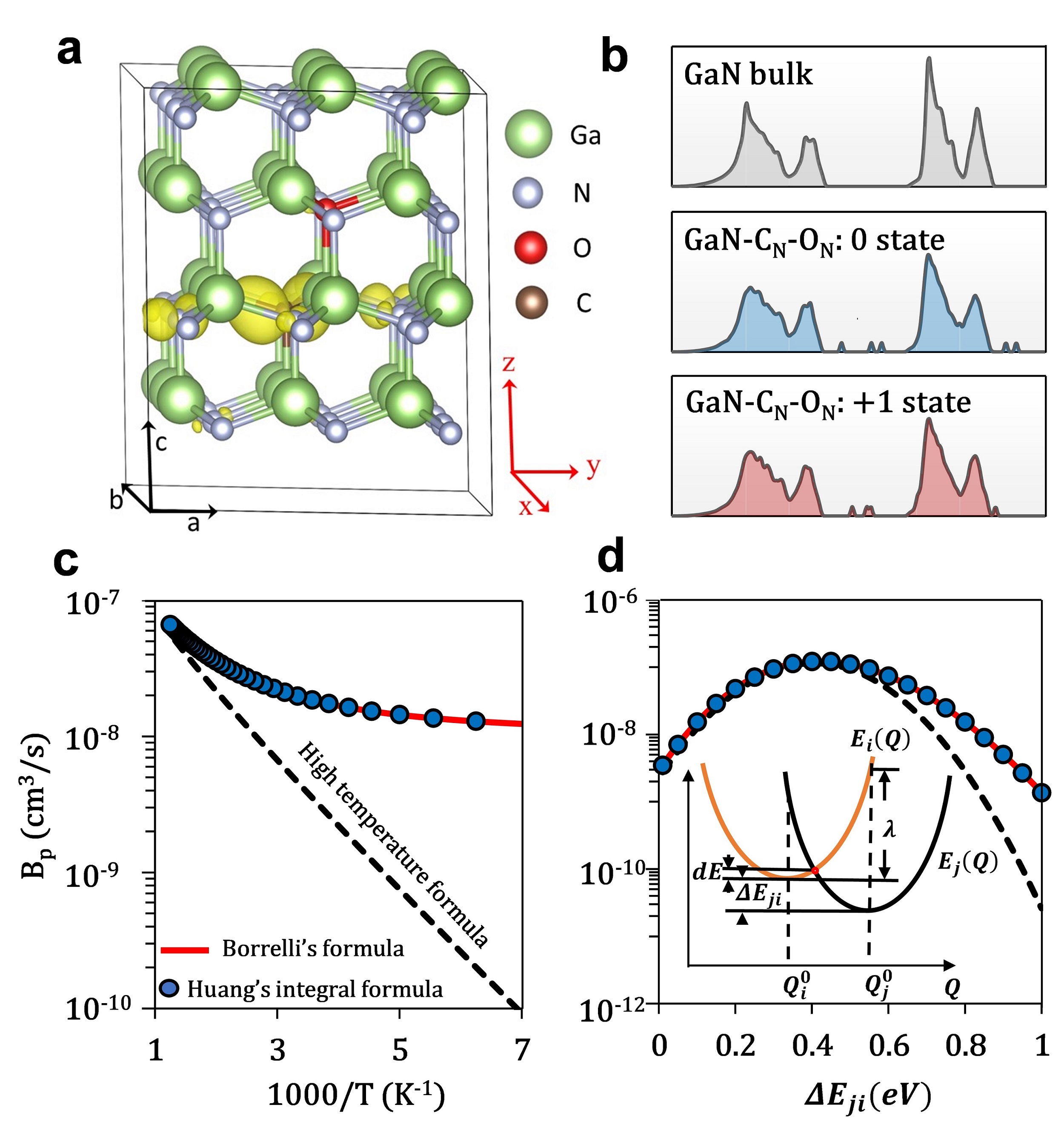}
\caption{\label{compare} (a)  The atom structure and wave function of the impurity state in the 72-atom supercell calculated using the HSE DFT functional. (b) The phonon spectrum for bulk GaN, neutral (0 state) and charged (+1 state) C$_N$+O$_N$ defects in 72-atom supercell. The shown phonon density of states (DOS) are calculated with PBE, but they are close to HSE results after a scaling as shown in Fig.S2. (c) The temperature-dependence of capture coefficients for different formulas; (d) The capture coefficients as function of transition energy $\Delta{E_{ji}}$ for different formulas at 300 K(same notation as in (c)). }
\end{figure}

The calculations of the defect transition energy have followed standard procedure (see SI-IV). Our HSE calculated transition energy is $E_{ji}$=0.77 eV above the VBM for the defect complex C$_N$+O$_N$, which agrees well with the experimental measured value of 0.85 eV \cite{wl31}. The reorganization energy $\lambda (\lambda_{(+)})$ is calculated by the energy relaxation of the charged (C$_N$+O$_N$$)^+$ defect starting from the equilibrium atomic coordinates $(Q_{ik}^0)$ of the neutral state $i$  under HSE functional. There is however another way to calculate this $\lambda$, to be called $\lambda_{(0)}$, which is the energy relaxation of the neutral C$_N$+O$_N$ defect starting from the equilibrium atom coordinates $(Q_{jk}^0)$ of the charged (C$_N$+O$_N$$)^+$ state. If the harmonic approximation is correct, these two $\lambda$ should be the same, both equals to $\lambda=\Sigma_k{(\omega_k^2/2)Q_{jik}^2}$. In reality, due to anharmonic, they are different. For the GaN:C$_N$+O$_N$ defect, $\lambda_{(0)}$ is 0.406 eV, while $\lambda_{(+)}$ is 0.518 eV. The HSE impurity state wave function is localized in the 72-atom supercell as shown in Fig.1(a). These wave functions are used in the variational method \cite{wl17} to calculate the electron-phonon coupling constant $\langle{\psi_i}|\partial{H}/\partial{R}|{\psi_j}\rangle$.

The calculated phonon modes and phonon DOSs are illustrated in Fig.1(b) (see SI-V for comparison between HSE and PBE results). We can see that the defect structure has a significant correction on the bulk phonon DOS. There are some new localized phonon mode peaks within the gap and near the edge of the optical band.

To compare with experiment, the  $W_{ij}$ of Eq.(1), (2) is converted to transition rate  $B_p=W_{ij}\cdot V_{sc}$, which is independent of super cell volume $V_{sc}$ \cite{wl36}. We first calculate $B_p$  based on neutral defect quantities. The temperature dependence results are shown in Fig.1(c), while in Fig.1(d) we also provided a result assuming the transition energy $\Delta{E_{ji}}$ could be different from our calculated values of 0.77 eV. We have also calculated the transition rate using the matrix based \emph{Borrelli}'s formula. As we can see, \emph{Huang}'s integral formula and \emph{Borrelli}'s formula give the exact same results, indicating that these two sets of formulas are equivalent (although they look rather different). As for \emph{Huang}'s high-T formula, it is good for temperature above 500 K, but significant difference appears for lower temperatures. However, this strongly depends on $\Delta{E_{ji}}$. For $\Delta{E_{ji}}<\lambda$, which is called the normal Marcus region, \emph{Huang}'s high-T formula is very good even at room temperature as shown in Fig.1(d). However, for $\Delta{E_{ji}}>\lambda$, the so-called inverse Marcus region, the error for \emph{Huang}'s high-T formula can be significant.

 In the above calculation, we obtain $B_p$ as $4.6\times10^{-9} cm^3/s$ and $1.98\times10^{-8} cm^3/s$ at 300K for high-T formula and quantum mechanical integration formula respectively. In contrast, the experimental value is in the range of $B_p = 3\times10^{-7} \sim 6\times10^{-7} cm^3/s$ \cite{wl32}. There is still one order of magnitude difference.

We now discuss the anharmonic effects and introduce ways to incorporate such effects. We first note the difference between $\lambda_{(0)}$ (0.406 eV) and $\lambda_{(+)}$ (0.518 eV) from direct HSE relaxation calculation. If everything is harmonic, they should both equal to  $\lambda^p=\Sigma_k{(\omega_k^2/2)Q_{jik}^2}$. This $\lambda^p$ also has two values, $\lambda_{(0)}^p$ (0.487 eV) and $\lambda_{(+)}^p$ (0.419 eV) using $\Delta{R_{ij}}$ and the phonon modes in (0) and (+) defect states. They are all different.  To correct this difference, we can rescale the phonon frequency as: $\omega_k'^{(0/+)}=\omega_k^{(0/+)}(\lambda_{(0/+)}/\lambda_{(0/+)}^p)^{1/2}$. After this rescaling, the $\lambda=\Sigma_k{(\omega_k^2/2)Q_{jik}^2}$ formula calculated reorganization energies will be the same as the direct atomic relaxation calculated ones. Besides  phonon frequency, another important representation of the anharmonic effect is in the phonon modes. We have carried out a dot product of phonon modes between (0) and (+) states. As shown in Fig.S1, there are significant differences. 

\begin{figure}[htbp]
	\centering
	\includegraphics[bb=0 0 1800 1800,width=3in]{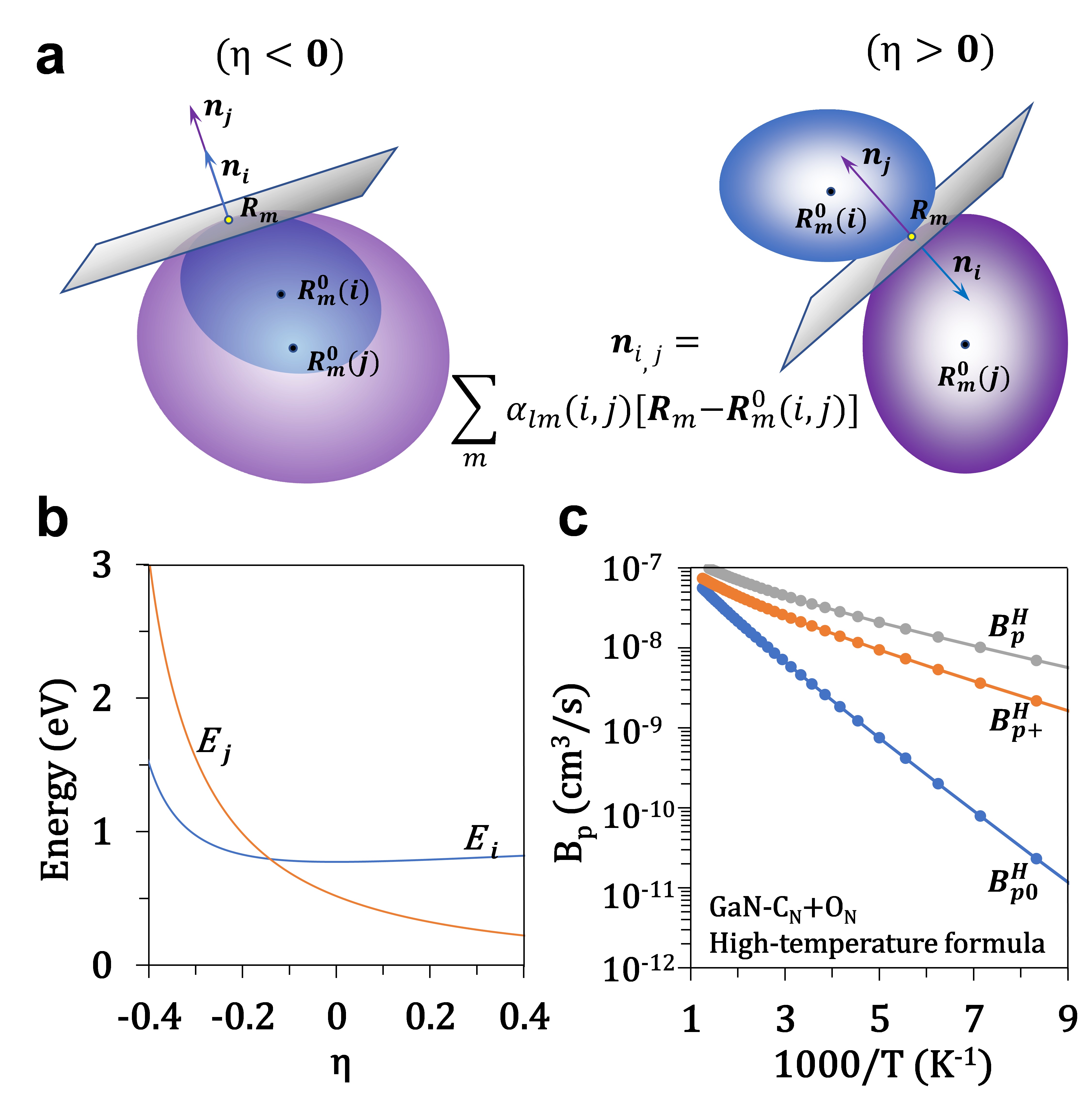}
	\caption{\label{compare} (a) The schematics of the configuration coordinate diagram for the defect. The purple and blue ellipsoid represent the adiabatic potentials for a defect in its positive and neutral charge state, respectively. (b) $E_{i(j)}$ as function of $\eta$. (c) The NMP transition rates $B_{p0}^H$ , $B_{p+}^H$ calculated from neutral charge state and + state phonon modes using Eq. (2), respectively. $B_p^H$ is the corrected result. }
\end{figure}

To take into account the anharmonic effects we first concentrate on the high-T formula of Eq.(2). The coupling constant $|V_C|^2=\Sigma_k{(1/K_BT{\omega_k^2})\langle{\psi_i}|\partial{H}/\partial{Q_k}|{\psi_j}\rangle^2}$ can be calculated separately using the renormalized phonon $\omega_k'$ from (0) and (+) states respectively, then a simple geometric average of these two $|V_C|^2$ together with the $1/\lambda^{1/2}$ prefactor in Eq.(2), can be used to get an average $|V_C|^2/\lambda^{1/2}$. Now, the exponential factor in Eq.(2) is just $exp(-dE_{ij}/k_BT)$, which is a thermal activation factor with $dE_{ij}$ being the barrier height (Fig.1(d)).  The barrier happens at the intersection of the two parabolic hyper surfaces of the two phonon systems. The minimum energy of such intersection point defines the barrier height $dE_{ij}$. If the two sets of phonon modes and frequencies at (0) and (+) are the same, then it is simple to show the minimum point happens along the straight line connecting the equilibrium positions of (0) and (+), and the barrier height $dE_{ij}$ equals $(\Delta{E_{ji}}-\lambda)^2/4\lambda$ as in Eq.(2) and Marcus formula. Now, one question is: if the phonon spectra of the two charged systems are different, how to calculate $dE_{ij}$.

We first note that, if an intersection point is the minimum energy point, then the two iso-energy elliptical surfaces of the two phonon systems must be tangent to each other. This is schematically shown in Fig.2 (a). That means the normal lines of these two elliptical surfaces at that point must be in the same orientation. The normal line equals the gradient of the phonon system energy at that point. This provides an important equation to help us to find the minimum point. Specifically, let's write the parabolic energy expression at the initial state $\emph{i}$(0):
\begin{equation}
 E_i (R)=E_i^0+\Sigma_{lm}{\alpha_{lm}(i)(R_l-R_l^0(i))(R_m-R_m^0 (i))},
 \end{equation}
where $\alpha_{lm}(i)=\Sigma_k{(\omega_k'^{(i)2}/2)M(l)\mu_k^{(i)}(l)\mu_k^{(i)}(m)M(m)}$, $\omega_k'^{(i)}$ is the renormalized phonon frequency, and  $E_i^0$, $R_m^0(i)$ denotes the equilibrium minimum energy and position for electronic state $i$ respectively and $M(l)$ is the atomic mass. The same expression applied to state $j(+)$.  Thus the condition for the gradients at point $\{$${R_m}$$\}$ to have the same orientation for state $i$ and $j$ can be expressed as (see SI-VI):
\begin{equation}
\Sigma_{lm}{\alpha_{lm}(i)(R_m-R_m^0(i))}=-\eta\Sigma_{lm}{\alpha_{lm}(j)(R_m-R_m^0(j))}.
\end{equation}
Here $\eta$ is a scaling parameter. Now, for a given $\eta$, we can solve the linear equation Eq.(4) for a solution $\{R_m (\eta)\}$, which can then be substitute into Eq.(3) to calculate the corresponding energies $E_j (\eta)$ and $E_i (\eta)$. The crossing of these two energies curves as a function of $\eta$ will give us the energy barrier as shown in Fig.2(b). Note in general cases when $\alpha_{lm}(i)\neq\alpha_{lm}(j)$, the crossing point is not on the straight line between $R_m^0(i)$ and $R_m^0(j)$. In our case of GaN:C$_N$+O$_N$, the so obtained barrier height is $dE'$=0.0218 eV. This can be compared with the barrier height $dE_0$=0.0833 eV, calculated using formula $(\Delta{E_{ji}}-\lambda(0))^2/4\lambda(0)$ from the neutral charge side and $dE_+$=0.0314 eV, using $\lambda_{(+)}$ from the + charge state. We see that, it is much better to use the final state reorganization energy (phonon spectrum) to estimate the barrier height if the original formula is going to be used. We have also calculated the 1D barrier height by seeking the intersection of the energy $E_i (R)$ of Eq.(3) and $E_j (R)$ along the straight line between the $R^0 (i)$ and $R^0 (j)$. The so obtained barrier height measured from $E_i^0$ is $dE(1D)$=0.0236 eV. In this case, it is only slightly larger than the true barrier height calculated above. To test this further, we have calculated the barrier heights as a function of $\Delta{E_{ji}}$, and the results are shown in Fig.S5(a),(b). We can see that the 1D formula can has significant error when $\Delta{E_{ji}}$ is large.

After we obtain the barrier height $dE'$  as above, we can use the activation factor $exp(-dE'/k_BT)$, and together with the above procedure for averaging $|V_C |^2/\lambda^{1/2}$  in Eq.(2) to calculate the transition rate. The so obtained transition rate $B_p^H$ is $3.914\times10^{-8} cm^3/s$ at 300 K, which is significantly larger than the transition rate obtained using the high-T formula from one set of phonon spectrum quantities (see Table I). The temperature dependence results are shown in Fig.2(c). 

\begin{figure}[htbp]
\centering
\includegraphics[bb=0 0 1800 1800,width=3in]{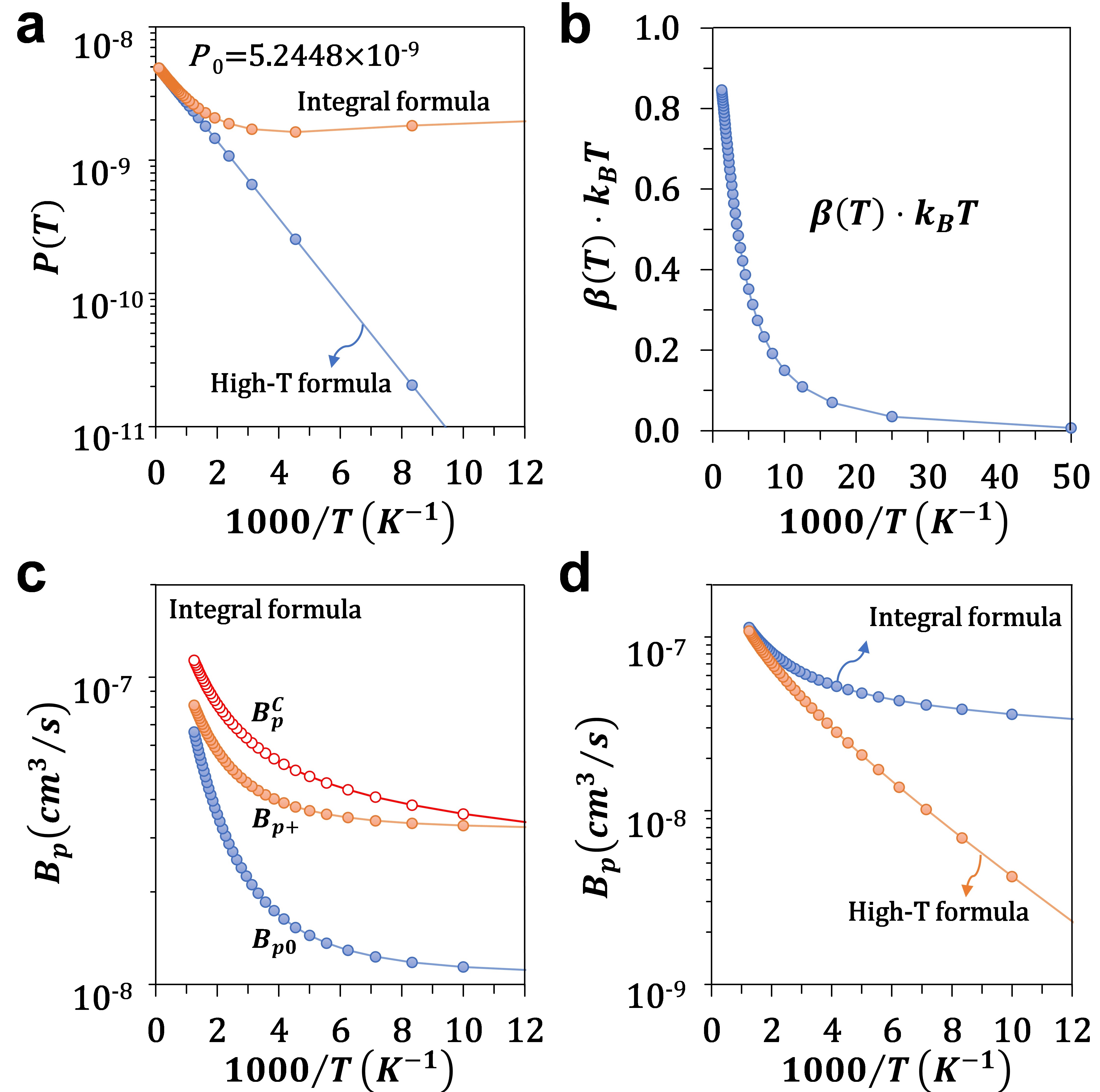}
\caption{\label{compare} (a) The function of the $P(T)$ with $1000/T$. (b) $\beta(T)\cdot k_BT$ as a function of $1000/T$. (c) The NMP translation rates calculated from different phonon modes as function of $1000/T$ for adiabatic integral approximation. (d) The corrected value of NMP translation rates as function of $1000/T$. }
\end{figure}

We next apply the above correction to quantum integration formula of Eq.(1). We note that the logarithmic of the quantum formula result does not follow the straight-line  $1/T$ relationship as shown in Fig.1(c). In order to apply our correction for the thermal activation energy, we can define a modified effective inverse temperature $\beta(T)$ to replace the $1/k_BT$. More specifically, we first define: $P(T)=(B_{p0}(T)B_{p+}(T))^{1/2}/{(T^{1/2})}$, and here $B_{p0}(T)$ and $B_{p+}(T)$ are the capture coefficients calculated using the corresponding normalized phonon frequencies $\omega_k'$  and phonon modes in the neutral (0) charge and (+) charge states by Eq.(1). We then assume $P(T)$ will have a temperature dependence as: $P(T)=P_0\times exp\{-\beta(T)(dE_0+dE_+)/2\}$. Here $dE_0$ and $dE_+$ are energy barriers calculated from the formula $(\Delta{E_{ji}}-\lambda)^2/4\lambda$ with the corresponding $\lambda_{(0/+)}$. At very large $T$, $\beta(T)=1/k_BT$, and one can obtain $ln(P_0)$ as the intersection of $\ln{P(T)}$ to $1/T=0$ axis as shown in Fig.3(a). Then, for a given $T$, $\beta(T)$ can be calculated as: $\beta(T)=-2 \ln{(P(T)/P_0)}/{(dE_0+dE_+)}$ as shown in Fig.3(b). After we get $\beta(T)$ at each $T$, we can then have our corrected temperature dependent quantum mechanical transition rate as:
 \begin{equation}
 B_p^C(T)=T^{1/2}P_0 e^{-\beta(T)dE'}.
 \end{equation}
Here the $dE'$ is the above calculated barrier height as shown in Fig.2(b) and Fig.S5(a). The $B_{p+}^C$ is shown in Fig.3(c), it is higher than both $B_{p0}$ and $B_{p+}$ using phonon parameters from neutral (0) charge and (+) charge states. Fig.3(d) shows a comparison between the integral formula and High-T formula both after the anharmonic corrections. At the room temperature, we yield resulting $B_p^C$ as $5.894\times10^{-8} cm^3/s$ and $3.914\times10^{-8}cm^3/s$ for quantum formula and high-T formula respectively. They are factors of 3 and 9 larger than the uncorrected results of $1.981\times10^{-8} cm^3/s$ and $4.608\times10^{-9}cm^3/s$, respectively.

The GaN:Zn$_{Ga}$-V$_N$ and 4H-SiC:V$_C$ defects are also calculated, with results summarized in Table.I at room temperature, while their temperature-dependences are shown in SI. We see that the anharmonic correction can increase the transition rate from 2 to 9 times. 

\begin{table}[htbp]
\caption{\label{compare} Calculated capture rates of different defects at 300K, compared with experiments. The units for the $B_p$ are $cm^3/s$. $B_{p0}^H$, $B_{p+}^H$ are the High-T formula using (0) and (+) charge phonon modes, respectively(Eq.(2)). $B_p^H$ is the anharmonic effect corrected High-T formula. $B_{p0}$ and $B_{p+}$ are the quantum integral formula (Eq.(1)) using (0) and (+) charge phonon modes,  respectively.  $B_p^C$ is the anharmonic effect corrected quantum formula result of Eq.(5).}

\begin{tabular}{cccccccc}
\hline
\hline
                &  GaN:C$_N$+O$_N$   & GaN:Zn$_{Ga}$-V$_N$  & 4H-SiC:V$_C$\\[1.0ex]\hline
$B_{p0}^H$      &   $4.61\times10^{-9}$   &  $2.48\times10^{-8}$  & $8.71\times10^{-8}$  \\
$B_{p+}^H$      &   $2.13\times10^{-8}$   &  $3.44\times10^{-8}$  & $8.88\times10^{-7}$  \\
$B_{p}^H$       &   $3.91\times10^{-8}$   &  $1.30\times10^{-7}$  & $8.63\times10^{-7}$  \\
$B_{p0}$        &   $1.98\times10^{-8}$   &  $5.13\times10^{-8}$  & $2.51\times10^{-7}$  \\
$B_{p+}$        &   $4.28\times10^{-8}$   &  $8.14\times10^{-8}$  & $9.95\times10^{-7}$  \\
$B_{p}^C$       &   $5.89\times10^{-8}$   &  $1.72\times10^{-7}$  & $1.05\times10^{-6}$  \\
$Expt.$         &   $6\times10^{-7}$ \cite{wl31}  &  $1\sim10\times10^{-7}$ \cite{wl38}  & over$1.8\times10^{-7}$ \cite{wl34}  \\
                &   $3.3\times10^{-8}$ \cite{wl37}   &  $3\sim30\times10^{-7}$ \cite{wl33}  &   \\
\hline
\hline
\end{tabular}
\end{table}

In summary, we have shown that the high-T NMP formula derived by \emph{Huang} provides an analytical expression for the coupling constant in Marcus theory for phonon induced coupling. But this coupling is different from that provided by a simple 1D model. We have also provided a practical procedure to calculate the correction due to anharmonic effect. Most importantly, a procedure is provided to calculate the energy barrier height between two different phonon spectra. We show that the overall anharmonic correction can increase the transition rate by a factor of 2 to 9.  The resulting nonradiative transition rates are closer to the experimental results. 

This work was partly supported by National Natural Science Foundation of China (Grand Nos. 61927901, 11674241, 11574304, 11774338), while L. W. W was supported by the Director, Office of Science (SC), Basic Energy Science (BES)/Materials Science and Engineering Division (MSED) of the U.S. Department of Energy (DOE) under the Contract No. DE-AC02-05CH11231 through the Theory of Material project. $^\ddagger$ these authors contribute equally to this paper.

\end{document}